\documentclass[aps,twocolumn,prb,amsmath,amssymb,showpacs,superscriptaddress]{revtex4}

\usepackage{graphicx,epsfig}

\newcommand{\re}{\mathrm{Re\,}}
\newcommand{\im}{\mathrm{Im\,}}
\newcommand{\ei}{\mathrm{Ei}}
\newcommand{\sgn}{\mathrm{sgn}}

\begin{document}
 
\title{Screening of a Luttinger liquid wire by a scanning tunneling microscope tip:\\
II. Transport properties}

\author{Marine Guigou}
\author{Thierry Martin}
\author{Adeline Cr\'epieux}

\affiliation{Centre de Physique Th\'eorique, Universit\'e de la M\'editerran\'ee, 163 avenue de Luminy, 13288 Marseille, France}

\date{\today}


\begin{abstract}
We study the effect of an electrostatic coupling between a scanning tunneling microscope tip and a Luttinger liquid wire on the tunneling current and noise between the two. Solving the Dyson equations non perturbatively for a local interaction potential, we derive the Green's functions associated to the wire and to the tip. Interestingly, the electrostatic coupling leads to the existence of new correlators, which we call mixed Green's functions, which are correlators between the bosonic fields of the wire and the tip. 
Next, we calculate the transport properties up to second order with the amplitude of the tunnel transfer: the tunnel current is strongly reduced by the presence of screening.  The zero-frequency noise is modified in a similar way, but the Fano factor remains unchanged. We also consider the effect of the screening on the asymmetry of the finite-frequency non-symmetrized noise and on the conductance.
\end{abstract}

\maketitle


\section{Introduction}

Transport properties in one-dimensional interacting systems such as quantum wires or metallic carbon nanotubes have been widely studied both theoretically and experimentally. 
When the wire or nanotube is well connected to a metallic contact, and another tunnel contact is placed on top of it, 
the current exhibits a zero bias anomaly \cite{kane,yao} because of the Coulomb interaction effects in this one dimensional system. On the contrary an isolated wire with bad contacts displays Coulomb blockade effects \cite{tans2}.  


Another interesting tool to characterize the transport is the noise (current-current fluctuations). It allows for instance to obtain information about the charge of the carriers flowing through the conductor. When electrons are injected in an {\it infinite length} carbon nanotube from a scanning tunneling microscope (STM) tip, it has been shown\cite{crepieux} that the Fano factor -- the ratio between the zero-frequency noise and the current -- is equal to $(K_c^2+1)e/2$, where $K_c$ is the Coulomb interaction strength. This Fano factor is tied to the charge of the collective excitations which are present within the wire. However, the presence of adiabatic Fermi liquid contacts at the extremities of the nanotube modifies\cite{lebedev} the Fano factor and effectively erases the information about the Coulomb interactions. 

To recover information about Coulomb interactions, finite-frequency noise is needed. Recent experiments\cite{deblock,billangeon} have shown that the current-current correlator which can be detected
is in general related to the non-symmetrized noise rather than the symmetrized one. Calculations of the finite-frequency non-symmetrized noise in the STM+nanotube system showed oscillations due to reflections at the contacts\cite{lebedev} whose period is related to the anomalous charges. 

In the above calculations, electrostatic effects such as screening between the wire and tip were typically left out.
Screening from the STM tip is likely to affect locally the electronic properties of the wire. Such effects were 
explicitly computed in Ref.~\onlinecite{guigou2}, where emphasis was put on the spectral properties of the wire.
There, we developed general expressions for the Dyson equations, which allowed to derive non perturbatively the 
Greens functions of the bosonic fields which are needed to obtain the spectral function of the wire. The tunneling density of states was shown to be enhanced (reduced) for large (weak) Coulomb interaction \cite{guigou2}. 

While the spectral function, the tunneling and the local density of states are quantities of great interest in condensed matter physics problems, they are typically measured in the context of electron transport, by tunneling experiments. 
In Ref.~\onlinecite{fisher_glazman} for instance, the tunneling current between two Luttinger liquids is typically 
expressed as a convolution of the two tunneling density of states of such materials. 

In the present paper, we wish for the first time to address transport and screening effects on the same footing. We argue that because the interaction between the tip and wire mixes the fermionic degrees of freedom of both source and drain, 
the derivation of the tunneling current and noise has to be revisited. As we shall see, the tunneling current depends on a new class of Greens function which we call ``mixed'' Green's functions. The use of the bosonization methods allows
to compute the tunneling current, as usual, to second order in the tunneling amplitude, albeit ``non perturbatively'' 
with respect to the Coulomb interaction between the wire and tip. This work thus suggests plainly that in the presence of 
screening effect, special care must be taken in the derivation of transport properties.   
A generalization of our results to carbon nanotubes can be straightforwardly obtained.

The structure of the paper is the following: in Sec.~II, we present the system and the Hamiltonian. In Sec.~III, we give the definitions of the tunnel current and of the non-symmetrized noise and their general expressions in term of bosonic Green's functions. In Sec.~IV, we calculate the mixed Green's functions. Next, in Sec.~V, we consider a particular form of the screening potential and we calculate the transport properties associated to an infinite length wire. We conclude in Sec.~VI.


\section{Model}

\begin{figure}[ht]
\begin{center}
\includegraphics[width=5cm]{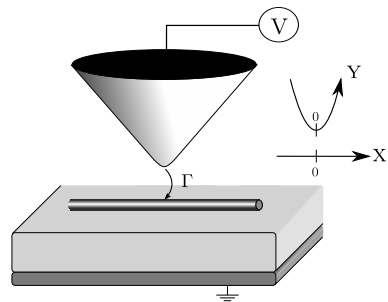}
\caption{Electron injection by a STM tip in a quantum wire. $V$ is the applied voltage to the tip and $\Gamma$ is the tunnel amplitude.\label{system}}
\end{center}
\end{figure}

The total Hamiltonian we consider is $H=H_0+H_V$ where $H_0$ is the unperturbed Hamiltonian and $H_V$ is the perturbation due to the voltage applied to the tip (see Fig.~\ref{system}). 

The unperturbed Hamiltonian is given by $H_0=H_W+H_T+H_{Sc}$, where $H_W$
describes the quantum wire:
\begin{eqnarray}
H_{W}&=&\sum_{j=c,s}\int_{-\infty}^{+\infty}dx \frac{v_{j}(x)}{2}\Big[K_{j}(x)\big(\partial_x\phi_{j}(x)\big)^2\nonumber\\
&&+K_{j}^{-1}(x)\big(\partial_x\theta_{j}(x)\big)^2\Big]~,
\label{H_W}
\end{eqnarray}

with $\theta_j$ and $\phi_j$ are the non-chiral bosonic fields, $K_c<1$ is the repulsive Coulomb interactions parameter in the charge sector, and $K_s=1$ is the Coulomb interactions parameter in the spin sector. $v_j=v_F/K_j$ is the velocity of the excitations in each sector $j$. We put $\hbar=1$ in the whole paper, excepted in voltage dependent terms and in figure captions.

$H_T$  describes the STM tip:
\begin{eqnarray}
H_{T}&=&\frac{u_F}{4\pi}\sum_{\sigma=\pm 1}\int_{-\infty}^{+\infty}dy~\big(\partial_y\varphi_{\sigma}(y)\big)^2~,
\label{H_T}
\end{eqnarray}
with $\varphi_{\sigma}$, the chiral bosonic field of the STM tip and $u_F$ is the Fermi velocity of electrons in the tip. $\sigma$ is the index which refers to the spin degree of freedom. The $x$ and $y$ axes are defined in Fig.~\ref{system}.

And, $H_{Sc}$ describes the screening:
\begin{eqnarray}
H_{Sc}=\int_{-\infty}^{+\infty} dx~ \int_{-\infty}^{+\infty} dy~\rho_{W}(x)W(x,y)\rho_{T}(y)~,
\label{H_Sc}
\end{eqnarray}

where $W$ is the electrostatic potential caused by the proximity between the STM tip and the quantum wire. $\rho_{W}$ is the density operator associated to the wire and $\rho_{T}$, the density operator associated to the tip:
\begin{eqnarray}
\rho_{W}(x)&=&\sqrt{\frac{2}{\pi}}\partial_x \theta_{\mathrm{c}}(x)+\rho_{2k_{\mathrm{F}}}(x)~,\\
\rho_{T}(y)&=&\frac{1}{2\pi}\sum_{\sigma}\partial_y\varphi_{\sigma}(y)~.
\end{eqnarray}

When $K_c>1/2$, we have shown in Ref.~\onlinecite{guigou2} that the $2k_F$--contribution to the density operator is not a relevant contribution to the screening Hamiltonian. We neglect it in this work.

The perturbation here is caused by a voltage $V$ applied to the tip: electrons can by injected from the tip to the wire or, inversely from the wire to the tip, by tunnel effect:
\begin{eqnarray}
H_V=\sum_{r=\pm,\sigma=\pm}\bigg[\Gamma(t)\psi^\dag_{r\sigma}(0)c_\sigma(0)+\Gamma^*(t)c^\dag_\sigma(0)\psi_{r\sigma}(0)\bigg]~,\nonumber\\
\end{eqnarray}

where $\Gamma(t)=\Gamma_0\exp(ieVt/\hbar)$ is the tunnel amplitude and the chirality $r$ refers to the right movers ($r=+$) or to the left movers ($r=-$). We assume that the electrons are injected in the wire at the position of the tip, i.e. at position $x=0$. $c^\dag_\sigma$ and $c_\sigma$ are the creation and annihilation operators attached to the tip:
\begin{eqnarray}
c_\sigma(y)&=&\frac{g_{\sigma}}{\sqrt{2\pi a}}e^{i\varphi_\sigma(y)}~,
\end{eqnarray}
 
where $g_{\sigma}$ is a Majorana fermion associated to the tip, and $a$ is the ultraviolet cut-off. $\psi^\dag_{r\sigma}$ and $\psi_{r\sigma}$ are the creation and annihilation operators attached to the chirality $r$ and to the spin $\sigma$ of the electrons in the wire:
\begin{eqnarray}
\psi_{r\sigma}(x)&=&\frac{f_{r\sigma}}{\sqrt{2\pi a}}e^{irk_Fx+i\sqrt{\frac{\pi}{2}}\sum_{r,\sigma}h_{\sigma}(j)(\phi_j(x)+r\theta_j(x))}~,\nonumber\\
\label{psi}
\end{eqnarray}

where $f_{r\sigma}$ is the Majorana fermion associated to the wire, $k_F$ is the Fermi wave vector. The coefficient $h_\sigma(j)$ is equal to $1$ when $j=c$ and to $\sigma$ when $j=s$.


\section{Transport properties}

In this section, we give the general expressions of the tunnel current and the non-symmetrized noise as a function of the bosonic Green's functions in the presence of screening.

\subsection{Tunnel current definition}

The average tunnel current is defined as:
\begin{eqnarray}
\langle I_T\rangle&=&\frac{1}{2}\sum_{\eta=\pm}\langle T_K\big\{I_T(t^{\eta})e^{-i\int_Kdt_1H_V(t_1)}\big\}\rangle~,
\end{eqnarray}

where $T_K$ denotes time ordering along the Keldysh contour\cite{keldysh}, $\eta$ indexes the position of time on it, and $\int_K dt_1\equiv\sum_{\eta_1=\pm}\int dt_1^{\eta_1}$. The tunnel current operator is defined by:
\begin{eqnarray}
I_T(t)&=&ie\sum_{r,\sigma}\bigg(\Gamma(t)\psi^\dag_{r\sigma}(0,t)c_\sigma(0,t)\nonumber\\
&&-\Gamma^*(t)c^\dag_\sigma(0,t)\psi_{r\sigma}(0,t)\bigg)~.
\end{eqnarray}

Up to the second order of the perturbative calculation with the tunnel amplitude $\Gamma_0$, the average tunnel current can be expressed as
\begin{widetext}
\begin{eqnarray}\label{tunnel_current}
\langle I_T\rangle&=&\frac{ie\Gamma_0^2}{(2\pi a)^2}\sum_{r,\sigma,\eta,\eta_1}\eta_1  \int_{-\infty}^{+\infty}d\tilde{t} \sin\left(\frac{eV\tilde{t}}{\hbar}\right)e^{\mathbf{G}_\sigma^{\varphi\varphi,\eta\eta_1}(0,0,\tilde{t})}\nonumber\\
&\times&e^{\frac{\pi}{2}\sum_j\big(\mathbf{G}^{\phi\phi,\eta\eta_1}_{j}(0,0,\tilde{t})+\mathbf{G}^{\theta\theta,\eta\eta_1}_{j}(0,0,\tilde{t})+r\mathbf{G}^{\phi\theta,\eta\eta_1}_{j}(0,0,\tilde{t})+r\mathbf{G}^{\theta\phi,\eta\eta_1}_{j}(0,0,\tilde{t})\big)}\nonumber\\
&\times&
e^{-\sqrt{\frac{\pi}{2}}\big(\mathbf{G}^{\phi\varphi,\eta\eta_1}_{c\sigma}(0,0,\tilde{t})+r\mathbf{G}^{\theta\varphi,\eta\eta_1}_{c\sigma}(0,0,\tilde{t})+\mathbf{G}^{\varphi\phi,\eta\eta_1}_{c\sigma}(0,0,\tilde{t})+r\mathbf{G}^{\varphi\theta,\eta\eta_1}_{c\sigma}(0,0,\tilde{t})\big)}~,
\end{eqnarray}
\end{widetext}

where $\tilde{t}\equiv t-t_1$. $\mathbf{G}^{\varphi\varphi,\eta\eta_1}_{\sigma}$ is the Keldysh Green's function of the tip in the presence of the electrostatic coupling defined as:
\begin{eqnarray}
&&{\bf G}_\sigma^{\varphi\varphi,\eta\eta_1}(0,0,\tilde{t})\equiv{\bf G}_\sigma^{\varphi\varphi,\eta\eta_1}(0,t^\eta;0,t_1^{\eta_1})\nonumber\\
&&=\langle\varphi_\sigma(0,t^\eta)\varphi_\sigma(0,t_1^{\eta_1})\rangle-\frac{1}{2}\langle\varphi_\sigma(0,t^\eta)\varphi_\sigma(0,t^\eta)\rangle\nonumber\\
&&-\frac{1}{2}\langle\varphi_\sigma(0,t_1^{\eta_1})\varphi_\sigma(0,t_1^{\eta_1})\rangle~.
\end{eqnarray}

Similary,  $\mathbf{G}^{\phi\phi,\eta\eta_1}_{j}$, $\mathbf{G}^{\theta\phi,\eta\eta_1}_{j}$,$\mathbf{G}^{\phi\theta,\eta\eta_1}_{j}$ and $\mathbf{G}^{\theta\theta,\eta\eta_1}_{j}$ are the Keldysh Green's functions of the wire in the presence of the electrostatic coupling. And finally, $\mathbf{G}^{\phi\varphi,\eta\eta_1}_{c\sigma}$, $\mathbf{G}^{\theta\varphi,\eta\eta_1}_{c\sigma}$,$\mathbf{G}^{\varphi\phi,\eta\eta_1}_{c\sigma}$ and $\mathbf{G}^{\varphi\theta,\eta\eta_1}_{c\sigma}$ are the mixed Keldysh Green's functions which are non-zero only in the presence of screening. 


Eq.~(\ref{tunnel_current}) differs to what one obtains in the absence of screening, in the sense that mixed Green's functions are not present in this case\cite{crepieux}. We call them mixed Green's functions, because they mix the bosonic fields $\theta$ or $\phi$ of the wire to the bosonic field $\varphi$ of the tip. The appearance of these new correlators is due to the electrostatic coupling between the wire and the tip, and they will affect the transport properties. Notice that these mixed Green's functions do not have any influence on the spectral properties of the wire because in the definitions of the spectral function and of the density of states, it only appears the bosonic fields of the wire\cite{guigou2}. On the contrary, in the transport properties, the bosonic fields of the wire and of the tip are mixed due to the term $H_V$ in the Hamiltonian, which corresponds to a tunnel transfer of electrons between the two sub-systems.

The calculation of the non perturbative Green's functions associated to $H_0$: $\mathbf{G}^{\theta\theta}_j$, $\mathbf{G}^{\phi\theta}_j$, $\mathbf{G}^{\theta\phi}_j$,$\mathbf{G}^{\phi\phi}_j$, and $\mathbf{G}^{\varphi\varphi}_\sigma$ is presented in detail in Ref.~\onlinecite{guigou2}, and the calculation of the mixed Green's functions is done in Sec.~IV.

\subsection{Non-symmetrized noise definition}

We define the non-symmetrized noise as
\begin{eqnarray}
S_T(t,t')&=&\langle T_K\big\{I_T(t^{-})I_T(t'^{+})e^{-i\int_Kdt_1H_V(t_1)}\big\}\rangle~.\nonumber\\
\end{eqnarray}

Up to the second order with the tunnel amplitude $\Gamma_0$, the non-symmetrized noise reads
\begin{widetext}
\begin{eqnarray}\label{tunnel_noise}
S_T(t,t')&=&\frac{e^2\Gamma_0^2}{(2\pi a)^2}\sum_{r,\sigma}\cos\left(\frac{eV(t-t')}{\hbar}\right)e^{ \mathbf{G}_\sigma^{\varphi\varphi,-+}(0,0,t-t')}\nonumber\\
&\times&e^{\frac{\pi}{2}\sum_j\big(\mathbf{G}^{\phi\phi,-+}_{j}(0,0,t-t')+\mathbf{G}^{\theta\theta,-+}_{j}(0,0,t-t')+r\mathbf{G}^{\phi\theta,-+}_{j}(0,0,t-t')+r\mathbf{G}^{\theta\phi,-+}_{j}(0,0,t-t')\big)}\nonumber\\
&\times&
e^{-\sqrt{\frac{\pi}{2}}\big(\mathbf{G}^{\phi\varphi,-+}_{c\sigma}(0,0,t-t')+r\mathbf{G}^{\theta\varphi,-+}_{c\sigma}(0,0,t-t')+\mathbf{G}^{\varphi\phi,-+}_{c\sigma}(0,0,t-t')+r\mathbf{G}^{\varphi\theta,-+}_{c\sigma}(0,0,t-t')\big)}~.
\end{eqnarray}
\end{widetext}

Since we consider an applied voltage constant in time, the noise depends only on the difference $\tilde{t}=t-t'$: $S_T(t,t')=S_T(t-t')$, we can thus define the following Fourier transform for the noise:
\begin{eqnarray}
S_T(\Omega)&=&\int_{-\infty}^{+\infty} d\tilde{t}e^{i\Omega\tilde{t}}S_T(\tilde{t})~.
\end{eqnarray}

In what follows, we will study both the zero-frequency noise $S_T(\Omega=0)$ and the finite-frequency noise $S_T(\Omega\ne 0)$. However, we need before to determine the Dyson equations of the mixed Green's functions which appear in the expressions of the current and the noise.


\section{Dyson equations for the mixed Green's functions}


The first step is to write the partition function in the Matsubara formalism\cite{QFT}:
\begin{eqnarray}
Z[\varphi,\phi,\theta]&=&\int {\mathcal D}\varphi_{\uparrow}{\mathcal D}\varphi_{\downarrow}{\mathcal D}\phi_c{\mathcal D}\phi_s{\mathcal D}\theta_c{\mathcal D}\theta_s e^{-\int d\tau (L_0-L_{aux})}~,\nonumber\\
\end{eqnarray}

where $L_0=L_W+L_T+L_{Sc}$ is the Lagrangian associated to $H_0$ and $L_{aux}$ contains auxiliary fields $\eta_{\theta_j}$, $\eta_{\phi_j}$ and $\eta_{\varphi_\sigma}$ needed to extract the Green's functions, such as for example ${\bf G}_{j\sigma}^{\varphi\theta}(y,\tau;x',\tau')=\langle T_{\tau}\{\varphi_{\sigma}(y,\tau)\theta_j(x',\tau')\}\rangle$, through the relation:
\begin{eqnarray}
\mathbf{G}_{j\sigma}^{\varphi\theta}(y,\tau;x',\tau')
=\frac{1}{Z}\frac{\partial^2Z}{\partial\eta_{\varphi_{\sigma}}(y,\tau)\partial\eta_{\theta_j}(x',\tau')}~,
\end{eqnarray}

where $T_{\tau}$ is the time ordering operator. As the Hamiltonian $H_0$ is quadratic with the bosonic fields $\theta_j$, $\phi_j$ and $\varphi_{\sigma}$, it is possible to calculate exactly the full Green's function by integrating out the degrees of freedom of the tip. We obtain
\begin{eqnarray}\label{mixed_dyson_1}
&&\mathbf{G}^{\varphi\theta}_{j\sigma}(y,\tau;x',\tau')=-\delta_{jc}\int d\tau_1\int_{-\infty}^{+\infty}dx_1\int_{-\infty}^{+\infty}dy_1\nonumber\\
&&\times \mathbf{G}^{\varphi\varphi}_{\sigma}(y,\tau;y_1,\tau_1)G^{-1}_{Sc}(x_1,y_1)G^{\theta\theta}_j(x_1,\tau_1;x',\tau')~,\nonumber\\
\end{eqnarray}

where $G^{-1}_{Sc}(x_1,y_1)=\partial_{x_1}\partial_{y_1}W(x_1,y_1)/(\pi\sqrt{2\pi})$, and $G^{\theta\theta}_j$ the Green's function of the wire without screening.

The derivation of the partition function with other auxiliary fields gives access, in a similar way, to the Dyson equations verified by the other mixed correlators:
\begin{eqnarray}\label{mixed_dyson_2}
&&\mathbf{G}^{\theta\varphi}_{j\sigma}(x,\tau;y',\tau')=-\delta_{jc}\int d\tau_1\int_{-\infty}^{+\infty}dx_1\int_{-\infty}^{+\infty}dy_1\nonumber\\
&&\times\mathbf{G}_{j}^{\theta\theta}(x,\tau;x_1,\tau_1)G^{-1}_{Sc}(x_1,y_1)G^{\varphi\varphi}_{\sigma}(y_1,\tau_1;y',\tau')~,\nonumber\\
\end{eqnarray}

\begin{eqnarray}\label{mixed_dyson_3}
&&\mathbf{G}^{\varphi\phi}_{j\sigma}(y,\tau;x',\tau')=-\delta_{jc}\int d\tau_1\int_{-\infty}^{+\infty}dx_1\int_{-\infty}^{+\infty}dy_1\nonumber\\
&&\times\mathbf{G}_{\sigma}^{\varphi\varphi}(y,\tau;y_1,\tau_1)G^{-1}_{Sc}(x_1,y_1)G^{\theta\phi}_j(x_1,\tau_1;x',\tau')~,\nonumber\\
\end{eqnarray}

and,
\begin{eqnarray}\label{mixed_dyson_4}
&&\mathbf{G}^{\phi\varphi}_{j\sigma}(x,\tau;y',\tau')=-\delta_{jc}\int d\tau_1\int_{-\infty}^{+\infty}dx_1\int_{-\infty}^{+\infty}dy_1\nonumber\\
&&\times\mathbf{G}_j^{\phi\theta}(x,\tau;x_1,\tau_1)G^{-1}_{Sc}(x_1,y_1)G^{\varphi\varphi}_{\sigma}(y_1,\tau_1;y',\tau')~.\nonumber\\
\end{eqnarray}

In the next section, we present the application of these results to an infinite length wire and we make the approximation, (which is justified in Ref. \onlinecite{guigou2}) that the double derivative of the electrostatic potential is represented by a local interaction.


\section{Application to a local electrostatic potential double derivative $\partial_x\partial_yW(x,y)$}

For the same reasons that are given in Ref.~\onlinecite{guigou2}, we assume an electrostatic potential of the form
\begin{eqnarray}
\partial_x\partial_yW(x,y)&=&W_0\delta(y)\delta(x)~.
\label{derive_potential}
\end{eqnarray}

With this specific form of the electrostatic potential and for an infinite length wire, we are able to solve the Dyson equations verified by the Green's functions associated to $H_0$.

The Keldysh Green's function associated to the bosonic field $\varphi$ of the tip that we need in Eqs.~(\ref{tunnel_current}) and (\ref{tunnel_noise}) reads\cite{guigou2}:
\begin{eqnarray}\label{green1}
&&\mathbf{G}^{\varphi\varphi,-+}_{\sigma}(0,0,\tilde{t})=G^{\varphi\varphi,-+}_{\sigma}(0,0,\tilde{t})+\frac{1}{2}\Bigg[2\ln(1+i\omega_c\tilde{t})\nonumber\\
&&-e^{\omega_{Sc}\tilde{t}-i\frac{\omega_{Sc}}{\omega_c}}\ei\left(-\omega_{Sc}\tilde{t}+i\frac{\omega_{Sc}}{\omega_c}\right)+e^{-i\frac{\omega_{Sc}}{\omega_c}}\ei\left(i\frac{\omega_{Sc}}{\omega_c}\right)\nonumber\\
&&-e^{-\omega_{Sc}\tilde{t}+i\frac{\omega_{Sc}}{\omega_c}}\ei\left(\omega_{Sc}\tilde{t}-i\frac{\omega_{Sc}}{\omega_c}\right)+e^{i\frac{\omega_{Sc}}{\omega_c}}\ei\left(-i\frac{\omega_{Sc}}{\omega_c}\right)\Bigg]~,\nonumber\\
\end{eqnarray}

where $\omega_c=v_F/a$ is the frequency cut-off, $\omega_{Sc}=(W_0/\pi)\sqrt{K_c/2}$ is the screening frequency, $\ei$ is the exponential integral function, and $G^{\varphi\varphi}_{\sigma}$ is the bare Green's function of the tip in the absence of screening. Notice that Eq.~(\ref{green1}) was derived exactly since no restriction has been imposed to the strength of the screening potential. The only assumption we have made is to take $K_c>1/2$ in order to neglect the $2k_F$-contribution in $\rho_W$.

The Keldysh Green's function attached to the bosonic field $\theta$ of the wire is:
\begin{eqnarray}\label{green2}
&&\mathbf{G}^{\theta\theta,-+}_{c}(0,0,\tilde{t})=G^{\theta\theta,-+}_{c}(0,0,\tilde{t})+\frac{K_c}{4\pi}\Bigg[2\ln(1+i\omega_c\tilde{t})\nonumber\\
&&-e^{\omega_{Sc}\tilde{t}-i\frac{\omega_{Sc}}{\omega_c}}\ei\left(-\omega_{Sc}\tilde{t}+i\frac{\omega_{Sc}}{\omega_c}\right)+e^{-i\frac{\omega_{Sc}}{\omega_c}}\ei\left(i\frac{\omega_{Sc}}{\omega_c}\right)\nonumber\\
&&-e^{-\omega_{Sc}\tilde{t}+i\frac{\omega_{Sc}}{\omega_c}}\ei\left(\omega_{Sc}\tilde{t}-i\frac{\omega_{Sc}}{\omega_c}\right)+e^{i\frac{\omega_{Sc}}{\omega_c}}\ei\left(-i\frac{\omega_{Sc}}{\omega_c}\right)\Bigg]~.\nonumber\\
\end{eqnarray}

For the spin sector ($j=s$), the Green's function is not affected by the screening: $\mathbf{G}^{\theta\theta,\eta\eta_1}_{s}(0,0,\tilde{t})=G^{\theta\theta,\eta\eta_1}_{s}(0,0,\tilde{t})$. The other Keldysh Green's functions associated to the wire are aslo unchanged:
\begin{eqnarray}
\mathbf{G}^{\theta\phi,\eta\eta_1}_{j}(0,0,\tilde{t})&=&G^{\theta\phi,\eta\eta_1}_{j}(0,0,\tilde{t})~,\\
\mathbf{G}^{\phi\theta,\eta\eta_1}_{j}(0,0,\tilde{t})&=&G^{\phi\theta,\eta\eta_1}_{j}(0,0,\tilde{t})~,\\
\mathbf{G}^{\phi\phi,\eta\eta_1}_{j}(0,0,\tilde{t})&=&G^{\phi\phi,\eta\eta_1}_{j}(0,0,\tilde{t})~.
\end{eqnarray}


The mixed Green's functions $\mathbf{G}^{\varphi\theta,-+}_{c\sigma}$ and $\mathbf{G}^{\theta\varphi,-+}_{c\sigma}$ are calculated in Appendix A, we obtain:
\begin{eqnarray}
&&\mathbf{G}^{\varphi\theta,-+}_{c\sigma}(0,0,\tilde{t})=\mathbf{G}^{\theta\varphi,-+}_{c\sigma}(0,0,\tilde{t})\nonumber\\
&&=\frac{1}{4}\sqrt{\frac{K_c}{2}}\sgn(1+i\omega_c\tilde{t})\sin\left[\frac{\omega_{Sc}}{\omega_c}(1+i\omega_c\tilde{t})\right]~.
\end{eqnarray}

The two other mixed Green's functions, which appear in Eqs.~(\ref{tunnel_current}) and (\ref{tunnel_noise}), are equal to zero even in the presence of screening:
\begin{eqnarray}
\mathbf{G}^{\phi\varphi,\eta\eta_1}_{j\sigma}(0,0,\tilde{t})&=&0~,\\
\mathbf{G}^{\varphi\phi,\eta\eta_1}_{j\sigma}(0,0,\tilde{t})&=&0~,
\end{eqnarray}

because the Hamiltonian $H_{Sc}$ couples only the $\theta$ and $\varphi$ fields.

Now, we have all the ingredients to calculate the transport properties of the wire in the presence of screening by the tip.

\begin{figure}[h!]
\begin{center}
\includegraphics[width=6.5cm]{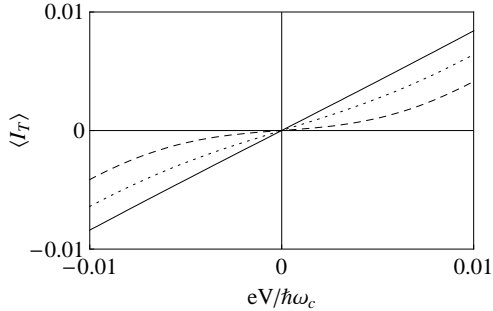}
\includegraphics[width=6.5cm]{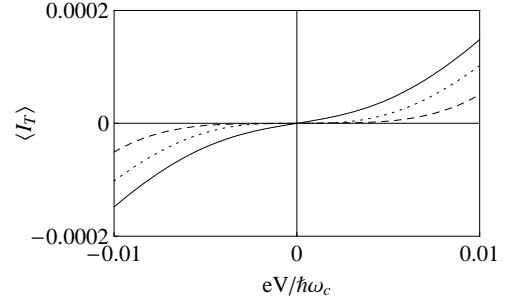}
\caption{Top panel: tunnel current as a function of voltage, in unit of $e\Gamma_0^2/(av_F\hbar^2\pi^3)$, for $K_c=0.7$. Bottom panel: tunnel current for $K_c=0.2$. On both graphics, the screening strength is $\omega_{Sc}=0$ (solid line), $\omega_{Sc}/\omega_c=0.001$ (dotted line), and $\omega_{Sc}/\omega_c=0.002$ (dashed line).}\label{figureCURRENT}
\end{center}
\end{figure}

\subsection{Tunnel current}

When we report these Green's functions in Eq.~(\ref{tunnel_current}), we obtain up to second order in $\Gamma_0$:
\begin{eqnarray}
\langle I_T\rangle&=&-\frac{2e\Gamma_0^2}{(2\pi a)^2}\sum_\sigma\int_{-\infty}^{+\infty}d\tilde{t} \sin\left(\frac{eV\tilde{t}}{\hbar}\right)\nonumber\\
&\times&\im\Bigg[\cosh\left(2\sqrt{\frac{\pi}{2}}\mathbf{G}^{\theta\varphi,-+}_{c\sigma}(0,0,\tilde{t})\right)e^{\mathbf{G}_\sigma^{\varphi\varphi,-+}(0,0,\tilde{t})}\nonumber\\
&\times&e^{\frac{\pi}{2}\sum_j\big(\mathbf{G}^{\phi\phi,-+}_{j}(0,0,\tilde{t})+\mathbf{G}^{\theta\theta,-+}_{j}(0,0,\tilde{t})\big)}\Bigg]~,
\end{eqnarray}

where we have performed the explicit sums over $r$ and over the Keldysh indices: only the $\eta\ne\eta_1$ terms contribute to the integral, because for $\eta=\eta_1$, the function we have to integrate is odd in $\tilde{t}$. After some algebra, we obtain:
\begin{widetext}
\begin{eqnarray}
\langle I_T\rangle&=&-\frac{2e\Gamma_0^2}{\pi^2a^2}\int_{0}^{+\infty}d\tilde{t} \sin\left(\frac{eV\tilde{t}}{\hbar}\right)\im\Bigg[\cosh\left(\frac{\sqrt{\pi K_c}}{4}\sgn(1+i\omega_c\tilde{t})\sin\left(\frac{\omega_{Sc}}{\omega_c}(1+i\omega_c\tilde{t})\right)\right)e^{-(\nu+1)\ln(1+i\omega_c\tilde{t})}\nonumber\\
&\times&
e^{\left(\frac{K_c}{8}+\frac{1}{2}\right)\left(2\ln(1+i\omega_c\tilde{t})-e^{\omega_{Sc}\tilde{t}-i\frac{\omega_{Sc}}{\omega_c}}\ei\left(-\omega_{Sc}\tilde{t}+i\frac{\omega_{Sc}}{\omega_c}\right)+e^{-i\frac{\omega_{Sc}}{\omega_c}}\ei\left(i\frac{\omega_{Sc}}{\omega_c}\right)-e^{-\omega_{Sc}\tilde{t}+i\frac{\omega_{Sc}}{\omega_c}}\ei\left(\omega_{Sc}\tilde{t}-i\frac{\omega_{Sc}}{\omega_c}\right)+e^{i\frac{\omega_{Sc}}{\omega_c}}\ei\left(-i\frac{\omega_{Sc}}{\omega_c}\right)\right)}\Bigg]~,\nonumber\\
\end{eqnarray}
\end{widetext}

where $\nu=(K_c+1/K_c+2)/4$, and where the summation over $\sigma$ has been made assuming a non-magnetic STM tip. To simplify the notations, we assume identical Fermi velocities in the wire and in the tip: $v_F=u_F$.

In Fig.~\ref{figureCURRENT} the tunneling current is plotted  as a function of applied voltage. For weak Coulomb interactions (top panel of Fig.~\ref{figureCURRENT}) and in the absence of screening, we see that the I-V characteristic is linear (solid line). When the tip is approached close to the wire, the I-V characteristic is modified: the amplitude of the current is reduced and the I-V curve follows a power law. For strong Coulomb interactions (bottom panel of Fig.~\ref{figureCURRENT}), the I-V characteristic follows a power law in the absence of screening. When the tip is approached close to the wire, the tunnel current stays a power law but is strongly reduced. 

The interpretation of these results is the following: electrostatic interactions between the tip and the wire plays a similar role than Coulomb interactions inside the wire: they both reduce the tunnel current and change the linear behavior of the I-V curve to a power law behavior. The tunnel current is strictly linear only in the absence of Coulomb interactions and electrostatic screening.


\subsection{Zero-frequency noise}

Up to second order with the tunnel amplitude $\Gamma_0$, the zero-frequency noise reads:
\begin{widetext}
\begin{eqnarray}
S_T(\Omega=0)&=&\frac{e^2\Gamma_0^2}{(2\pi a)^2}\sum_{r,\sigma}\int_{-\infty}^{+\infty} d\tilde{t}\cos\left(\frac{eV\tilde{t}}{\hbar}\right)e^{2\pi \mathbf{G}_\sigma^{\varphi\varphi,-+}(0,0,\tilde{t})}\nonumber\\
&\times&e^{\frac{\pi}{2}\sum_j\big(\mathbf{G}^{\phi\phi,-+}_{j}(0,0,\tilde{t})+\mathbf{G}^{\theta\theta,-+}_{j}(0,0,\tilde{t})+r\mathbf{G}^{\phi\theta,-+}_{j}(0,0,\tilde{t})+r\mathbf{G}^{\theta\phi,-+}_{j}(0,0,\tilde{t})\big)}\nonumber\\
&\times&
e^{-\sqrt{\frac{\pi}{2}}\big(\mathbf{G}^{\phi\varphi,-+}_{c\sigma}(0,0,\tilde{t})+r\mathbf{G}^{\theta\varphi,-+}_{c\sigma}(0,0,\tilde{t})+\mathbf{G}^{\varphi\phi,-+}_{c\sigma}(0,0,\tilde{t})+r\mathbf{G}^{\varphi\theta,-+}_{c\sigma}(0,0,\tilde{t})\big)}~.
\end{eqnarray}
\end{widetext}

\begin{figure}[h!]
\begin{center}
\includegraphics[width=6.5cm]{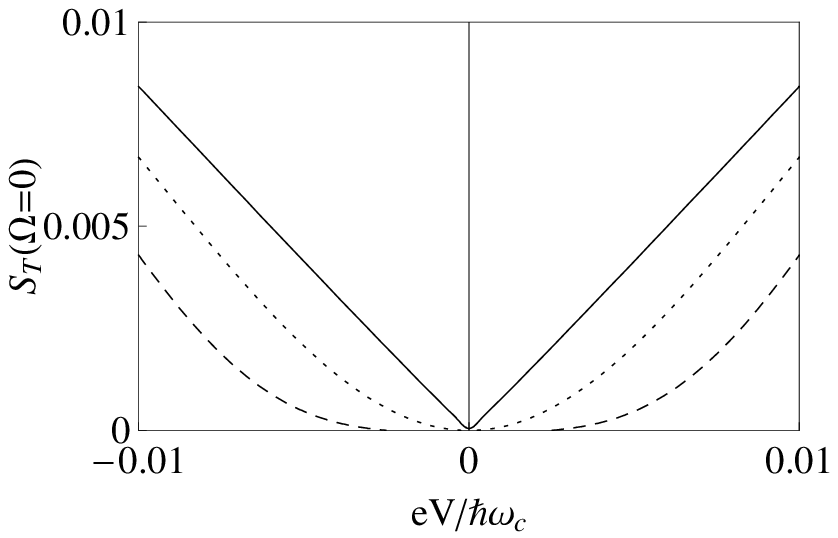}
\includegraphics[width=6.5cm]{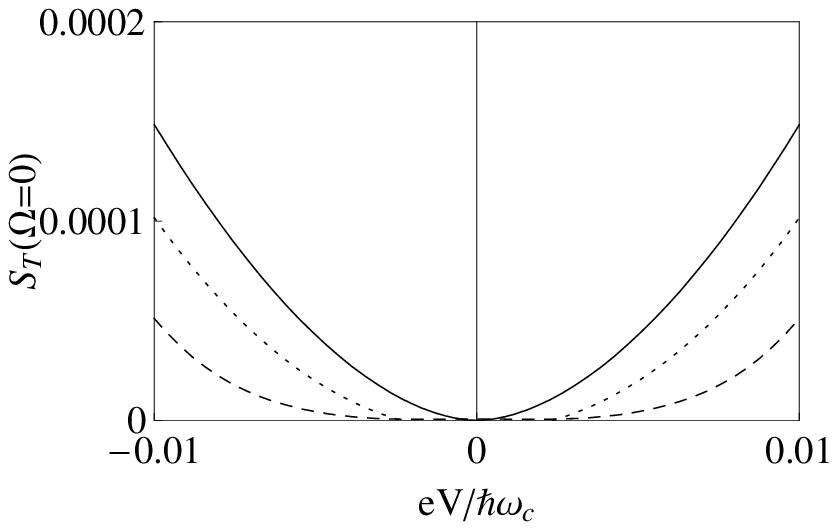}
\caption{Top panel: zero-frequency noise as a function of voltage, in unit of $e^2\Gamma_0^2/(av_F\hbar^2\pi^3)$, for $K_c=0.7$. Bottom panel: zero-frequency noise for $K_c=0.2$. On both graphics, the screening strength is $\omega_{Sc}=0$ (solid line), $\omega_{Sc}/\omega_c=0.001$ (dotted line), and $\omega_{Sc}/\omega_c=0.002$ (dashed line).}\label{figureNOISE}
\end{center}
\end{figure}

When one reports the expressions of the Green's functions, it reduces to:
\begin{eqnarray}
&&S_T(\Omega=0)=\frac{e^2\Gamma_0^2}{2\pi^2a^2}\sum_{\sigma}\int_{-\infty}^{+\infty} d\tilde{t}\cos\left(\frac{eV\tilde{t}}{\hbar}\right)\nonumber\\
&&\times \cosh\Big(2\sqrt{\frac{\pi}{2}}\mathbf{G}^{\theta\varphi,-+}_{c\sigma}(0,0,\tilde{t})\Big)e^{2\pi \mathbf{G}_\sigma^{\varphi\varphi,-+}(0,0,\tilde{t})}\nonumber\\
&\times&e^{\frac{\pi}{2}\sum_j\big(\mathbf{G}^{\phi\phi,-+}_{j}(0,0,\tilde{t})+\mathbf{G}^{\theta\theta,-+}_{j}(0,0,\tilde{t})\big)}~,
\end{eqnarray}

which leads to
\begin{widetext}
\begin{eqnarray}
S_T(\Omega=0)&=&\frac{2e^2\Gamma_0^2}{\pi^2a^2}\int_{0}^{+\infty}d\tilde{t} \cos\left(\frac{eV\tilde{t}}{\hbar}\right)\re\Bigg[\cosh\left(\frac{\sqrt{\pi K_c}}{4}\sgn(1+i\omega_c\tilde{t})\sin\left(\frac{\omega_{Sc}}{\omega_c}(1+i\omega_c\tilde{t})\right)\right)e^{-(\nu+1)\ln(1+i\omega_c\tilde{t})}\nonumber\\
&\times&
e^{\left(\frac{K_c}{8}+\frac{1}{2}\right)\left(2\ln(1+i\omega_c\tilde{t})-e^{\omega_{Sc}\tilde{t}-i\frac{\omega_{Sc}}{\omega_c}}\ei\left(-\omega_{Sc}\tilde{t}+i\frac{\omega_{Sc}}{\omega_c}\right)+e^{-i\frac{\omega_{Sc}}{\omega_c}}\ei\left(i\frac{\omega_{Sc}}{\omega_c}\right)-e^{-\omega_{Sc}\tilde{t}+i\frac{\omega_{Sc}}{\omega_c}}\ei\left(\omega_{Sc}\tilde{t}-i\frac{\omega_{Sc}}{\omega_c}\right)+e^{i\frac{\omega_{Sc}}{\omega_c}}\ei\left(-i\frac{\omega_{Sc}}{\omega_c}\right)\right)}\Bigg]~,\nonumber\\
\end{eqnarray}
\end{widetext}

where the summation over $\sigma$ has been made assuming a non-magnetic STM tip.

In the top panel of Fig.~\ref{figureNOISE}, we plot the zero-frequency noise as a function of applied voltage for weak Coulomb interactions. The zero-frequency noise, which is linear in the absence of screening (solid line), follows a power law in the presence of screening, similarly to what we obtain for the tunnel current. For strong Coulomb interactions, the zero-frequency noise follows a power law whatever the screening strength is (bottom panel of Fig.~\ref{figureNOISE}). The effect of the screening is to reduce strongly the zero-frequency noise.

An interesting quantity to consider is the Fano factor defined as the ratio between the zero-frequency noise and the average current:
\begin{eqnarray}
F&\equiv&\frac{S_T(\Omega=0)}{|\langle I_T\rangle|}~.
\end{eqnarray}

This Fano factor is related to the charge of the carrier which is transferred trough the conductor: for example, $F$ equals $e$ for a junction between normal metals, whereas $F$ equals $2e$ for a junction between a normal metal and a superconductor\cite{supra}. In a two-dimensional electron gas in the fractional quantum Hall regime, simultaneous measurements of the current and noise at a constriction were performed and a Fano factor equals to $e/3$, which corresponds to the fractional charge, was obtained\cite{saminadayar,picciotto}.

Here, the Fano factor stays equal to the electron charge $e$ even in the presence of screening. The reason is that only electron can tunnel from the tip to the wire and we end up with the standart Schottky relation\cite{schottky}:
\begin{eqnarray}
S_T(\Omega=0)=e|\langle I_T\rangle|~.
\end{eqnarray}

At zero frequency, the non-symmetrized noise and the symmetrized noise are identical because of the time translation invariance. It is not more the case when one studies the finite-frequency noise\cite{bena1,bena2}. In the next section, we focus our interest to the finite-frequency non-symmetrized noise $S_T(\Omega)$, knowing that the symmetric noise can be deduced simply by taking $(S_T(\Omega)+S_T(-\Omega))/2$.

\subsection{Finite-frequency non-symmetrized noise}

Up to the second order with the tunnel amplitude $\Gamma_0$, the finite-frequency non-symmetrized noise is:
\begin{widetext}
\begin{eqnarray}
S_T(\Omega)&=&\frac{e^2\Gamma_0^2}{(2\pi a)^2}\sum_{r,\sigma}\int_{-\infty}^{+\infty} d\tilde{t}e^{i\Omega\tilde{t}}\cos\left(\frac{eV\tilde{t}}{\hbar}\right)e^{2\pi \mathbf{G}_\sigma^{\varphi\varphi,-+}(0,0,\tilde{t})}\nonumber\\
&\times&e^{\frac{\pi}{2}\sum_j\big(\mathbf{G}^{\phi\phi,-+}_{j}(0,0,\tilde{t})+\mathbf{G}^{\theta\theta,-+}_{j}(0,0,\tilde{t})+r\mathbf{G}^{\phi\theta,\eta\eta_1}_{j}(0,0,\tilde{t})+r\mathbf{G}^{\theta\phi,-+}_{j}(0,0,\tilde{t})\big)}\nonumber\\
&\times&
e^{-\sqrt{\frac{\pi}{2}}\big(\mathbf{G}^{\phi\varphi,-+}_{c\sigma}(0,0,\tilde{t})+r\mathbf{G}^{\theta\varphi,-+}_{c\sigma}(0,0,\tilde{t})+\mathbf{G}^{\varphi\phi,-+}_{c\sigma}(0,0,\tilde{t})+r\mathbf{G}^{\varphi\theta,-+}_{c\sigma}(0,0,\tilde{t})\big)}~.
\end{eqnarray}
\end{widetext}

\begin{figure}[h!]
\begin{center}
\includegraphics[width=7cm]{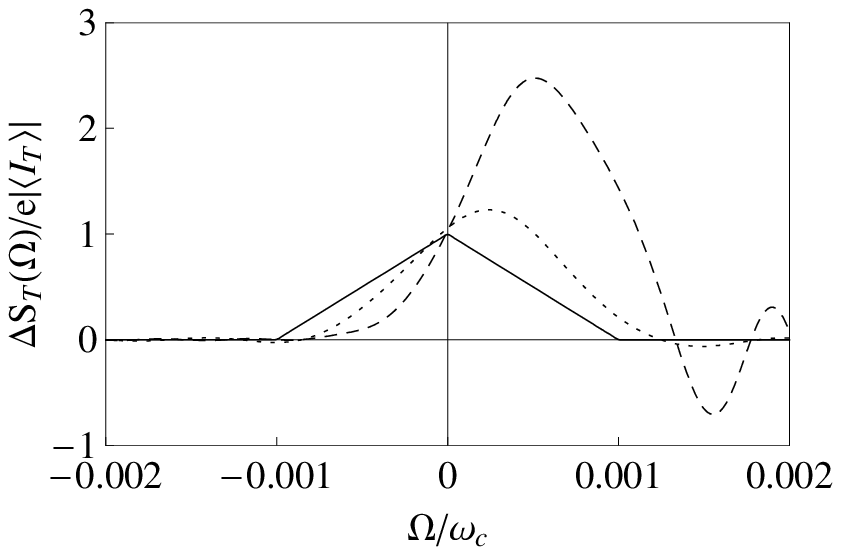}
\includegraphics[width=7cm]{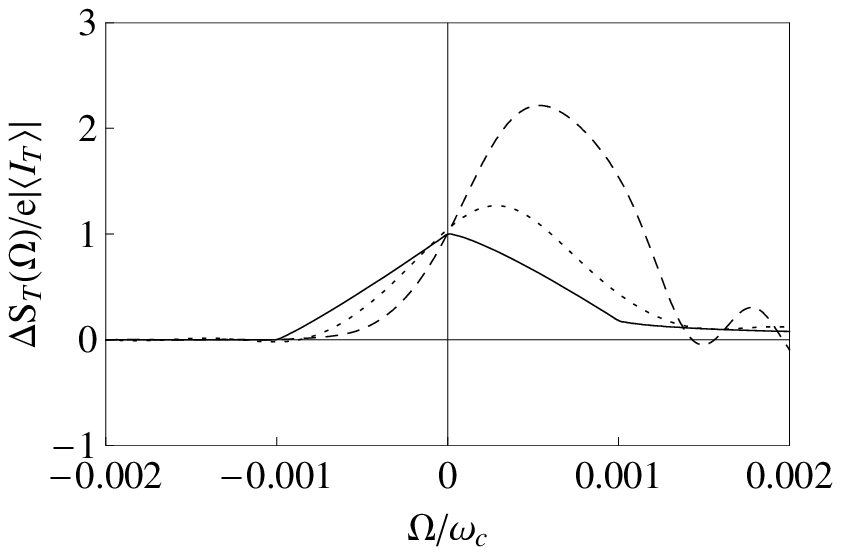}
\caption{Top panel: non-symmetrized excess noise as a function of frequency, divided by the tunnel current times the electron charge, in the absence of Coulomb interactions ($K_c=1$). Bottom panel: the same quantity in the presence of Coulomb interactions ($K_c=0.5$). On both graphics, the screening strength is $\omega_{Sc}/\omega_c=0$ (solid line), $\omega_{Sc}/\omega_c=0.0001$ (dotted line), and $\omega_{Sc}/\omega_c=0.0002$ (dashed line), and the voltage value is $eV/\hbar\omega_c=0.001$.}\label{figureFFNOISE}
\end{center}
\end{figure}

When one reports the expressions of the Green's functions, it reduces to:
\begin{eqnarray}
&&S_T(\Omega)=\frac{e^2\Gamma_0^2}{2\pi^2a^2}\sum_{\sigma}\int_{-\infty}^{+\infty} d\tilde{t}e^{i\Omega\tilde{t}}\cos\left(\frac{eV\tilde{t}}{\hbar}\right)\nonumber\\
&&\times \cosh\Bigg(2\sqrt{\frac{\pi}{2}}\mathbf{G}^{\theta\varphi,-+}_{c\sigma}(0,0,\tilde{t})\Bigg)e^{2\pi \mathbf{G}_\sigma^{\varphi\varphi,-+}(0,0,\tilde{t})}\nonumber\\
&\times&e^{\frac{\pi}{2}\sum_j\big(\mathbf{G}^{\phi\phi,-+}_{j}(0,0,\tilde{t})+\mathbf{G}^{\theta\theta,-+}_{j}(0,0,\tilde{t})\big)}~.
\end{eqnarray}

Finally,
\begin{widetext}
\begin{eqnarray}
S_T(\Omega)&=&\frac{2e^2\Gamma_0^2}{\pi^2a^2}\int_{0}^{+\infty}d\tilde{t} \cos\left(\frac{eV\tilde{t}}{\hbar}\right)\re\Bigg[\cosh\left(\frac{\sqrt{\pi K_c}}{4}\sgn(1+i\omega_c\tilde{t})\sin\left(\frac{\omega_{Sc}}{\omega_c}(1+i\omega_c\tilde{t})\right)\right)e^{i\Omega\tilde{t}-(\nu+1)\ln(1+i\omega_c\tilde{t})}\nonumber\\
&\times&
e^{\left(\frac{K_c}{8}+\frac{1}{2}\right)\left(2\ln(1+i\omega_c\tilde{t})-e^{\omega_{Sc}\tilde{t}-i\frac{\omega_{Sc}}{\omega_c}}\ei\left(-\omega_{Sc}\tilde{t}+i\frac{\omega_{Sc}}{\omega_c}\right)+e^{-i\frac{\omega_{Sc}}{\omega_c}}\ei\left(i\frac{\omega_{Sc}}{\omega_c}\right)-e^{-\omega_{Sc}\tilde{t}+i\frac{\omega_{Sc}}{\omega_c}}\ei\left(\omega_{Sc}\tilde{t}-i\frac{\omega_{Sc}}{\omega_c}\right)+e^{i\frac{\omega_{Sc}}{\omega_c}}\ei\left(-i\frac{\omega_{Sc}}{\omega_c}\right)\right)}\Bigg]~.\nonumber\\
\end{eqnarray}
\end{widetext}

In experiments, we have access to the total noise which contains many contributions, such as zero-point fluctuations, $1/f$ noise\cite{dutta} or thermal fluctuations\cite{johnson,nyquist}. If one wants to isolate the contribution due to the applied voltage (shot noise), one has to look at the non-symmetrized excess noise defined as:
\begin{eqnarray}
\Delta S_T(\Omega)&\equiv&S_T(\Omega)-S_T(\Omega)|_{V=0}~.
\end{eqnarray}

In the top panel of Fig.~\ref{figureFFNOISE}, we plot the non-symmetrized excess noise, divided by the tunnel current times the electron charge, in the absence of Coulomb interactions. In the unscreened case, the emission noise ($\Omega<0$) cancels when $\Omega<-eV/\hbar=-0.001$, and the absorption noise ($\Omega>0$) cancels when $\Omega>eV/\hbar=0.001$ (solid line in the top panel of Fig.~\ref{figureFFNOISE}). In the screened case, the emission noise cancels when $\Omega<-eV/\hbar$, whereas the absorption noise does not cancel any more when $\Omega>eV/\hbar$ (dotted and dashed lines in the top panel of Fig.~\ref{figureFFNOISE}).

In the presence of Coulomb interactions (bottom panel of Fig.~\ref{figureFFNOISE}), the non-symmetrized excess noise is always asymmetric, whatever the value of the screening strength is. From these graphics, we conclude that the Coulomb interactions inside the wire and the electrostatic potential due to the tip have a similar effect on the asymmetry of the finite-frequency excess noise. Moreover, the excess noise is symmetric only when both Coulomb interactions in the wire and screening by the STM tip are turned off.

\subsection{Conductance}

In Refs.~\onlinecite{bena2} and \onlinecite{safi2}, it has been shown that the difference between the absorption noise and the emission noise is related to the real part of the conductance:
\begin{eqnarray}
S_T(\Omega)-S_T(-\Omega)=2\hbar\Omega\re[G(\Omega)]~.
\end{eqnarray}

As a consequence, the real part of the excess conductance can easily be obtained by taking the difference between the excess noise at positive frequency and the excess noise at negative frequency:
\begin{eqnarray}
\re[\Delta G(\Omega)]=\frac{\Delta S_T(\Omega)-\Delta S_T(-\Omega)}{2\hbar\Omega}~,
\end{eqnarray}

where $\Omega>0$. 

\begin{figure}[h!]
\begin{center}
\includegraphics[width=7cm]{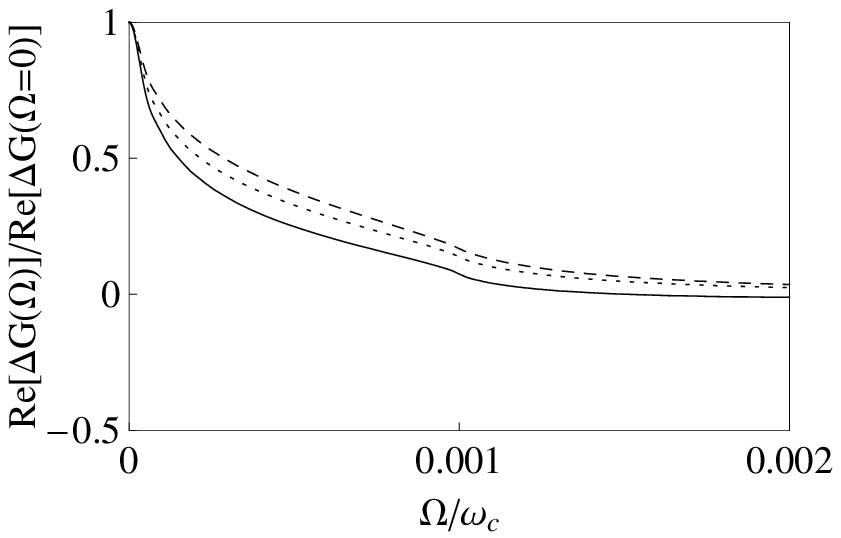}
\includegraphics[width=7cm]{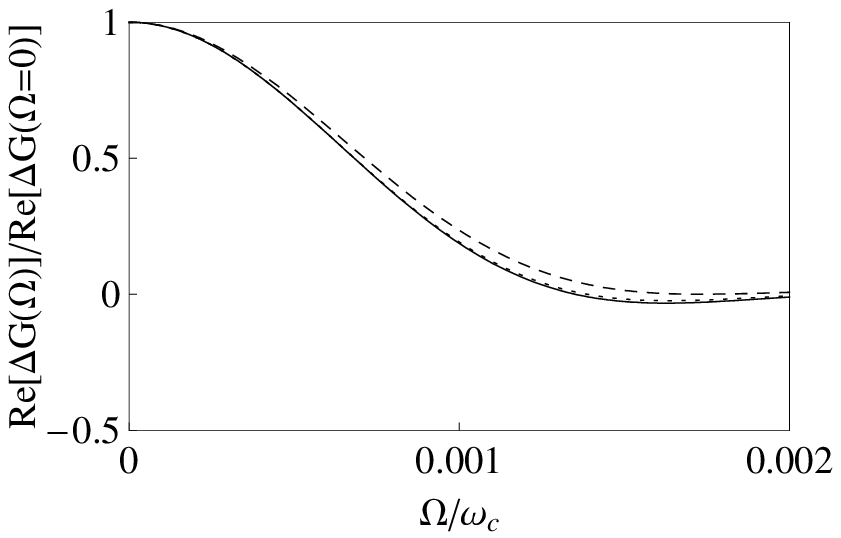}
\caption{Top panel: real part of the excess conductance as a function of frequency, normalized by its value at zero frequency, in the absence of screening.
Bottom panel: the same quantity in the presence of screening ($\omega_{Sc}/\omega_c=0.0002$). On both graphics, the Coulomb interactions is $K_c=0.9$ (solid line), $K_c=0.7$ (dotted line), and $K_c=0.5$ (dashed line), and the voltage value is $eV/\hbar\omega_c=0.001$.}\label{figureCOND}
\end{center}
\end{figure}

In Fig.~\ref{figureCOND}, we plot the real part of the excess conductance normalized by its value at zero frequency. The top panel of the Fig.~\ref{figureCOND} is obtained when screening is absent: we see that the excess conductance decreases rapidly and converges through its value in the absence of Coulomb interactions, which is zero. Indeed, in the absence of Coulomb interactions (i.e., for $K_c=1$), the current is linear with the voltage and as a consequence, the conductance is constant in voltage. Thus, the excess conductance is simply zero in this case. We also see a cusp when the frequency equals the voltage, here at $\Omega/\omega_c=eV/\hbar\omega_c=0.001$, which is attenuated when the Coulomb interactions strength increases. 

The bottom panel of the Fig.~\ref{figureCOND} shows the excess conductance in the presence of screening. We observe a weaker decreasing at low frequency in comparison to the unscreened case. More interestingly, the excess conductance becomes almost independent of the Coulomb interactions strength. This is due to the fact that the effect of screening by the STM tip dominates over the effect of the Coulomb interactions in the quantum wire.


\section{Conclusion}

In this paper, we have calculated the transport properties of a Luttinger liquid wire in the presence of electrostatic screening by a STM tip. Whereas the spectral properties of the wire depend only on the Green's functions of the wire\cite{guigou2}, the transport properties depend also on the Green's function of the tip and on new correlators that we call mixed Green's functions because they mix the bosonic fields of the wire with the bosonic field of the tip. These new correlators are non-zero due to the electrostatic tip-wire interaction. We have solved the Dyson equations associated to them in the case where the double derivative of the electrostatic potential can be approximated as a local interaction and we have calculated the tunnel current, the noise and the conductance.

We have shown that the tunnel current and the zero-frequency noise are strongly reduced in comparison to the unscreened situation. In addition, the I-V characteristic, which is linear in the absence of Coulomb interactions in the wire, follows a power law behavior when screening is turned on. We can conclude that the electrostatic interactions between the wire and the tip have a similar effect on the I-V curve than the electron-electron interactions inside the wire.

Next, we have studied the finite-frequency non-symmetrized excess noise and we obtain an interesting result. Whereas in the absence of screening, the non-symmetrized noise associated to a non-interacting wire is symmetric in frequency, it becomes asymmetric in the presence of screening. For an interacting wire, the non-symmetrized excess noise is always asymmetric in frequency regardless of the screening strength is. The explanation of this asymmetric signal for a non-interacting wire is again the fact that the electrostatic potential between the wire and the tip has similar effects as electron-electron interactions. The subtle difference here is that the effective local density-density interaction which is induced by the presence of the tip contains retardation effects: they involve the dynamics of the tip (via its Green's function).  

A useful tool to test the asymmetry of the excess noise is to calculate the excess conductance: indeed, when the excess noise is symmetric in frequency, the excess conductance is simply zero. We have calculated the excess conductance in different situations. In the absence of Coulomb interactions and screening, this quantity cancels. On the contrary, in the presence of Coulomb interactions, or of screening, the excess conductance does not cancel any more.

The results that we have obtained in this paper are preliminaries since we assume an infinite length wire. However, with the help of the Dyson equations that we have derived, it is now possible to calculate the transport properties for more realistic situations: in particular, in the presence of electrical contacts, at the extremities of the wire, which are needed to measure transport properties. We can estimate that current and noise will be affected by the contacts only for short length wire, due to finite size effects. Moreover, it would be interesting to calculate the current and its fluctuations in the wire, and not only the tunnel current and tunnel noise. It has been shown that the electrical contacts induce oscillating behavior of the finite-frequency noise in the wire due to reflections at the contacts\cite{guigou1,pugnetti}. The question of how these oscillations are modified by the screening, in the case where the electrostatic potential between the wire and the tip has a finite length extension, would also be interesting to consider.

\section{acknowledgments}

We thank D.~Bercioux, F.~Dolcini, H.~Grabert, F.~Hekking and I.~Safi for valuable discussions.


\appendix

\section{Expressions of the mixed Green's functions in the presence of screening}

The starting point is the Dyson equation given by Eq.~(\ref{mixed_dyson_1}) that verifies the mixed Green's function $\mathbf{G}^{\varphi\theta}_{j\sigma}$ in imaginary time:
\begin{eqnarray}
&&\mathbf{G}^{\varphi\theta}_{j\sigma}(y,\tau;x',\tau')=-\delta_{jc}\int d\tau_1\int_{-\infty}^{+\infty}dx_1\int_{-\infty}^{+\infty}dy_1\nonumber\\
&&\times\mathbf{G}_{\sigma}^{\varphi\varphi}(y,\tau;y_1,\tau_1)G^{-1}_{Sc}(x_1,y_1)G^{\theta\theta}_j(x_1,\tau_1;x',\tau')~,\nonumber\\
\end{eqnarray}

where $\mathbf{G}_\sigma^{\varphi\varphi}$ is the Green's function of the tip in the presence of screening, $G^{-1}_{Sc}(x_1,y_1)=W_0\delta(x_1)\delta(y_1)/(\pi\sqrt{2\pi})$, and $G_j^{\theta\theta}$ is the Green's function of the wire in the absence of screening. Since we assume that the tunnel transfert takes place at $x'=y=0$, we only need $\mathbf{G}^{\varphi\theta}_{j\sigma}(0,\tau;0,\tau')$ which obeys to:
\begin{eqnarray}
\mathbf{G}^{\varphi\theta}_{j\sigma}(0,\tau;0,\tau')&=&-\frac{\delta_{jc}W_0}{\pi\sqrt{2\pi}}\int d\tau_1\mathbf{G}_{\sigma}^{\varphi\varphi}(0,\tau;0,\tau_1)\nonumber\\
&&\times G^{\theta\theta}_j(0,\tau_1;0,\tau')~.
\end{eqnarray}

Using the time translation invariance, we perform a Fourier transform and obtain:
\begin{eqnarray}
\mathbf{G}^{\varphi\theta}_{j\sigma}(0,0,\omega)&=&-\frac{\delta_{jc}W_0}{\pi\sqrt{2\pi}}\mathbf{G}_{\sigma}^{\varphi\varphi}(0,0,\omega)G^{\theta\theta}_j(0,0,\omega)~.\nonumber\\
\end{eqnarray}

Reporting the expression of $\mathbf{G}_{\sigma}^{\varphi\varphi}$ which was determined in Ref.~\onlinecite{guigou2}:
\begin{eqnarray}
\mathbf{G}_{\sigma}^{\varphi\varphi}(0,0,\omega)=\frac{\pi|\omega|}{\omega^2-\omega^2_{Sc}}~,
\end{eqnarray}

we obtain
\begin{eqnarray}
\mathbf{G}^{\varphi\theta}_{j\sigma}(0,0,\omega)&=&-\delta_{jc}\frac{\sqrt{\pi K_c}}{2}\frac{\omega_{Sc}}{\omega^2-\omega^2_{Sc}}~,
\end{eqnarray}

whose Fourier transform is:
\begin{eqnarray}
\mathbf{G}^{\varphi\theta}_{j\sigma}(0,0,\tilde{\tau})&=&\frac{\delta_{jc}}{4}\sqrt{\frac{K_c}{2}}\sgn(\tilde{\tau})\sin(\omega_{Sc}\tilde{\tau})~.
\end{eqnarray}

Finally, performing an analytic continuation $\tilde{\tau}\rightarrow i\tilde{t}+\tau_0$ where $\tau_0=1/\omega_c$, we get the Keldysh mixed Green's function
\begin{eqnarray}
\mathbf{G}^{\varphi\theta,-+}_{j\sigma}(0,0,\tilde{t})&=&\frac{\delta_{jc}}{4}\sqrt{\frac{K_c}{2}}\sgn(i\tilde{t}+\tau_0)\sin(\omega_{Sc}(i\tilde{t}+\tau_0))~.\nonumber\\
\end{eqnarray}

Next, we consider the Dyson equation verified by $\mathbf{G}^{\theta\varphi}_{j\sigma}$ which is given by Eq.~(\ref{mixed_dyson_2}):
\begin{eqnarray}
&&\mathbf{G}^{\theta\varphi}_{j\sigma}(x,\tau;y',\tau')=-\delta_{jc}\int d\tau_1\int_{-\infty}^{+\infty}dx_1\int_{-\infty}^{+\infty}dy_1\nonumber\\
&&\times\mathbf{G}_{j}^{\theta\theta}(x,\tau;x_1,\tau_1)G^{-1}_{Sc}(x_1,y_1)G^{\varphi\varphi}_{\sigma}(y_1,\tau_1;y',\tau')~.\nonumber\\
\end{eqnarray}

The resolution of the equation is similar to the previous one, it leads to:
\begin{eqnarray}
\mathbf{G}^{\theta\varphi,-+}_{j\sigma}(0,0,\tilde{t})&=&\frac{\delta_{jc}}{4}\sqrt{\frac{K_c}{2}}\sgn(i\tilde{t}+\tau_0)\sin(\omega_{Sc}(i\tilde{t}+\tau_0))~.\nonumber\\
\end{eqnarray}

Now, we look at the Dyson equation verified by $\mathbf{G}^{\varphi\phi}_{j\sigma}$ which is given by Eq.~(\ref{mixed_dyson_3}):
\begin{eqnarray}
&&\mathbf{G}^{\varphi\phi}_{j\sigma}(y,\tau;x',\tau')=-\delta_{jc}\int d\tau_1\int_{-\infty}^{+\infty}dx_1\int_{-\infty}^{+\infty}dy_1\nonumber\\
&&\times\mathbf{G}_{\sigma}^{\varphi\varphi}(y,\tau;y_1,\tau_1)G^{-1}_{Sc}(x_1,y_1)G^{\theta\phi}_j(x_1,\tau_1;x',\tau')~.\nonumber\\
\end{eqnarray}

At positions $x'=y=0$, it reduces to:
\begin{eqnarray}
\mathbf{G}^{\varphi\phi}_{j\sigma}(0,\tau;0,\tau')&=&-\frac{\delta_{jc}W_0}{\pi\sqrt{2\pi}}\int d\tau_1\mathbf{G}_{\sigma}^{\varphi\varphi}(0,\tau;0,\tau_1)\nonumber\\
&&\times G^{\theta\phi}_j(0,\tau_1;0,\tau')~.
\end{eqnarray}

Since we have $G^{\theta\phi}_j(0,0,\omega)=0$ (see Appendix B), it leads to $G^{\theta\phi}_j(0,\tau_1;0,\tau')=0$ and as a consequence:
\begin{eqnarray}
\mathbf{G}^{\varphi\phi}_{j\sigma}(0,\tau;0,\tau')&=&0~.
\end{eqnarray}

Finally, we look at the Dyson equation for $\mathbf{G}^{\phi\varphi}_{j\sigma}$ which is given by Eq.~(\ref{mixed_dyson_4}):
\begin{eqnarray}
&&\mathbf{G}^{\phi\varphi}_{j\sigma}(x,\tau;y',\tau')=-\delta_{jc}\int d\tau_1\int_{-\infty}^{+\infty}dx_1\int_{-\infty}^{+\infty}dy_1\nonumber\\
&&\times\mathbf{G}_j^{\phi\theta}(x,\tau;x_1,\tau_1)G^{-1}_{Sc}(x_1,y_1)G^{\varphi\varphi}_{\sigma}(y_1,\tau_1;y',\tau')~.\nonumber\\
\end{eqnarray}

At positions $x'=y=0$, it reduces to:
\begin{eqnarray}
\mathbf{G}^{\phi\varphi}_{j\sigma}(0,\tau;0,\tau')&=&-\frac{\delta_{jc}W_0}{\pi\sqrt{2\pi}}\int d\tau_1\mathbf{G}_j^{\phi\theta}(0,\tau;0,\tau_1)\nonumber\\
&&\times G^{\varphi\varphi}_{\sigma}(0,\tau_1;0,\tau')~.
\end{eqnarray}

Since we have $G^{\phi\theta}_j(0,0,\omega)=0$ (see Appendix B), it leads to $G^{\phi\theta}_j(0,\tau_1;0,\tau')=0$ and as a consequence:
\begin{eqnarray}
\mathbf{G}^{\phi\varphi}_{j\sigma}(0,\tau;0,\tau')&=&0~.
\end{eqnarray}


\section{Expressions of the bare Green's functions}

For an infinite length wire and in the absence of screening, the Matsubara bosonic Green's functions at zero temperature and at positions $x=x'=0$ are:
\begin{eqnarray}
G^{\phi\phi}_{j}(0,0,\omega)&=&\frac{1}{2|\omega|K_{j}}~,\\
G^{\theta\theta}_{j}(0,0,\omega)&=&\frac{K_{j}}{2|\omega|}~,\\
G^{\theta\phi}_{j}(0,0,\omega)&=&G^{\phi\theta,0}_{j}(0,0,\omega)=0~.
\end{eqnarray}

The bosonic Green's function of the tip that we need in the calculation is the one located at $y=y'=0$ which is given at zero temperature by:
\begin{eqnarray}
G_{\sigma}^{\varphi\varphi}(0,0,\omega)&=&\frac{\pi}{|\omega|}~.
\end{eqnarray}


\end{document}